\documentclass[aps,prc,superscriptaddress,reprint,nofootinbib]{revtex4-1}

\usepackage{epsfig}
\usepackage{dcolumn}
\usepackage{bm}
\usepackage{amssymb}
\usepackage{amsmath}
\usepackage{hyperref}
\usepackage{comment}
\usepackage{ulem}

\usepackage[dvipsnames]{xcolor}
\usepackage{pgfplots}
\usepackage{graphicx}
\graphicspath{{./plots/}}
\pgfplotsset{compat=1.8}
\definecolor{darkblue}{rgb}{0.0, 0.0, 0.55}
\definecolor{cite}{rgb}{0.0, 0.34, 0.25}
\definecolor{midgreen}{rgb}{0.52, 0.73, 0.4}

\usepackage{enumitem}
\usepackage{rotating}
\usepackage{bigints}
\begin{document}
\title{Partial Pressure Contributions of Hadron Families to the QCD Equation of State}

\author{J. Gonzales}
\email[Corresponding author: ]{jagonz84@central.uh.edu}
\affiliation{Department of Physics, University of Houston, Houston, TX 77204, USA}

\author{A. Florez}
\affiliation{Department of Physics, University of Houston, Houston, TX 77204, USA}

\author{J. Jahan}
\affiliation{Department of Physics, University of Houston, Houston, TX 77204, USA}

\author{A. R. Nava Acuna}
\affiliation{Department of Physics, University of Houston, Houston, TX 77204, USA}

\author{N. Mehndiratta}
\affiliation{Department of Physics and Astronomy, UCLA, Los Angeles, CA 90095 USA}

\author{C. Ratti}
\affiliation{Department of Physics, University of Houston, Houston, TX 77204, USA}

\date{\today}

\begin{abstract}
Lattice simulations provide the thermodynamics of quantum chromodynamics (QCD) as a function of the temperature, at zero-to-moderate values of the baryonic chemical potential. However,  the contribution of single hadronic species cannot be directly isolated from lattice calculations. In this work, we find linear combinations of up to fourth order susceptibilities which isolate the contribution of hadrons to the QCD pressure according to their baryon number $B$, electric charge $Q$ and strangeness $S$ content. These combinations are valid, provided that the thermodynamics of a strongly-interacting gas in the low-temperature regime can be modeled as a gas of non-interacting hadrons and their resonances.
Finally, we test the validity of these linear combinations in the Hadron Resonance Gas (HRG) model and compare them to available lattice QCD results, using continuum-estimated susceptibilities.
\end{abstract}

\maketitle

\section{Introduction}
\label{sec:intro}
Quantum Chromodynamics (QCD) is the quantum field theory which describes the interaction between quarks and gluons. This non-Abelian, SU(3) gauge theory is formulated from only a handful of ingredients, yet it exhibits many interesting features---chiral symmetry breaking, asymptotic freedom, confinement, phase transitions, etc. \cite{Ratti:2022qgf,Ratti:2018ksb,Bzdak:2019pkr,Du:2024wjm,Sorensen:2023zkk}. 
On the experimental side, mapping out the QCD phase diagram on the $(T,\mu_B)$ plane of temperature and baryon chemical potential is an active field of research probed through relativistic heavy-ion collisions \cite{STAR:2010vob, ALICE:2022wpn} and, more recently, through astrophysical observations such as neutron star mergers \cite{LIGOScientific:2018cki, LIGOScientific:2020aai, Annala:2017llu}. 
Of utmost importance in this endeavor is understanding the deconfinement phase transition of QCD, \textit{i.e.} the temperature where the relevant degrees of freedom transition from bound-states of hadrons at low $T$, to a plasma of quarks and gluons at high $T$ \cite{Aoki:2006we, Fischer:2009wc}. 
Theoretical approaches, such as first principles lattice QCD \cite{Wilson:1974sk, Kogut:1974ag, Susskind:1976jm} or models like the Hadron Resonance Gas (HRG) \cite{Hagedorn:1965st, Kapusta:1982qd, Gorenstein:1981fa} supplement the experimental endeavor to understand the phase diagram of the strong force. The HRG model provides accurate results at low $T$, as well as a useful interpretation of the physics in that range. 
Lattice QCD simulations are the state-of-the-art method to achieve first-principle calculations from QCD. Observables such as the Equation of State (EoS) can be calculated through Monte-Carlo simulations of quark-gluon configurations on a discretized space-time hyper-lattice \cite{Karsch:2001cy, deForcrand:2002hgr, DElia:2002tig}. 
Both approaches allow us to obtain thermodynamic quantities such as the pressure, and although they agree at low temperatures, each of them has limitations. 
The HRG model cannot describe deconfinement, while lattice QCD does not allow us to isolate the specific contributions from different hadron families at low $T$, and is usually limited to temperature $T\geq 120$ MeV and small chemical potential.

In this manuscript, we follow the approach presented in Refs.
\cite{Bazavov:2013dta,Alba:2017mqu} to isolate the partial pressures of the different hadron families based on their quantum numbers, and to express them in terms of conserved charge susceptibilities. We improve the results of Ref. \cite{Alba:2017mqu} in two ways: for the first time, we are able to also separate the partial pressure contributions in terms of electric charge, besides strangeness and baryon number. This allows e.g. to distinguish between the proton and its resonances and the neutron and its resonances. Besides, we utilize linear combinations based on state-of-the-art continuum estimated susceptibilities \cite{Abuali:2025tbd, jahan_2025_16749145}, which allows a quantitative comparison with the HRG model results as well as a useful reference for model-builders, who are seeking to constrain their hadronic interaction parameters.
The code developed to calculate the partial pressures based on combinations determined in this work has been integrated into the MUSES calculation engine \cite{jahan_2026_19747641, calculation_engine_all_versions}.

The manuscript is organized as follows. After an introduction to the HRG model in Sec.~\ref{sec:hrg} and an explanation of how the concept of partial pressures can emerge from this model, we introduce the methodology used to obtain the partial pressures and present selected results in Sec.~\ref{sec:lincomb}, before summarizing our work in Sec.~\ref{sec:concl}.

\section{The Hadron Resonance Gas Model}
\label{sec:hrg}
The HRG model is a highly successful approach for describing strongly-interacting matter in the hadronic phase and it plays a crucial role in our formulation of partial pressures. Introduced in 1965 by Rolf Hagedorn \cite{Hagedorn:1965st}, this model assumes hadrons to be the fundamental degrees of freedom in a thermodynamic system. Starting from a system of hadrons in their ground state, the HRG model describes their interaction by creating heavier and heavier, non-interacting resonances. This non-kinetic usage of energy --- which increases the number of species and their masses, rather than the momentum per particle --- reproduces experimental observations. 
Furthermore, the model is known to have a limiting temperature, the so-called Hagedorn temperature $T_H$, which was interpreted by Parisi and Cabibbo as the transition temperature at which the system changes to different degrees of freedom, namely deconfined quarks and gluons forming the quark-gluon plasma \cite{CABIBBO197567}. The modern version of the HRG model partition function is summed over all known hadron species and resonances  categorized by the Particle Data Group (PDG) \cite{ParticleDataGroup:2024cfk}, and has the following form:
\begin{multline}
\label{hrgeq}
    \ln{Z}^{HRG}(T,V,{\mu})= \sum_{i\in PDG} \ln{Z}_i(T,V,{\mu})= \quad \\ 
    \quad \frac{Vg_i}{2\pi^2}\int_0^\infty \pm p^2 dp \ln\left[1\pm \lambda_i (T,{\mu})\exp(-\beta\epsilon_i)\right]
\end{multline}

where, for each particle species $i$:
\begin{itemize}
    \item + is for fermions and -- is for bosons
    \item $\epsilon_i=\sqrt{\textbf{p}^2+m_i^2}$ is the relativistic energy
    \item $g_i$ is the spin degeneracy factor
    \item $\lambda_i(T,{\mu})=\exp\left[\frac{B_i\mu_B+S_i\mu_S+Q_i\mu_Q}{T}\right]$ is the fugacity, with $\mu_B,~\mu_S$ and $\mu_Q$ the chemical potentials for baryon number, strangeness and electric charge and $B_i,~S_i$ and $Q_i$ the baryon number, strangeness and electric charge carried by hadron $i$.
\end{itemize}

By performing an expansion of the logarithm in Eq.~\eqref{hrgeq}, one can re-write the partition function as:
\begin{equation}
\label{hrglogexpeq}
    \ln{Z_i}(T,V,{\mu})=\frac{Vg_i}{2\pi^2}\sum^\infty_{N=1}\frac{(\pm1)^{N+1}\lambda_i^N}{N}\int p^2dp [\exp(-\beta\epsilon_i)]^N
\end{equation}
and then perform the integral to get:
\begin{equation}
\label{hrgboltzmann}
\ln{Z_i}(T,V,{\mu})=\frac{Vg_i}{2\pi^2}\sum^\infty_{N=1}\frac{(\pm1)^{N+1}\lambda_i^N}{N^2} m_i^2 K_2\left(N\frac{m_i}{T}\right),
\end{equation}
where $K_2$ is the Modified Bessel function of the second kind. By making use of the fugacity $\lambda_i$, one can express the partition function by pairing contributions from both particle and antiparticle states to make use of the $\cosh$ functions:
\begin{equation}
\begin{split}
\ln{\mathcal{Z}_i}(T,V,{\mu})=\frac{Vg_i}{2\pi^2}\sum^\infty_{N=1}\frac{(\pm1)^{N+1}}{N^2} m_i^2 K_2\left(N\frac{m_i}{T}\right)\\\times\cosh\left(N(B_i\mu_B+S_i\mu_S+Q_i\mu_Q)/T\right) \; . 
\end{split}
\label{pp}
\end{equation}
This form allows to isolate different hadron contributions to the full pressure, since one can distinguish them via their quantum numbers, namely baryon number $B$, strangeness $S$ and electric charge $Q$. Keeping the first term in this expansion ($N=1$) corresponds to the Boltzmann approximation, where higher terms are suppressed as long as $(m_N-\mu_B) \geq T$ \cite{Vovchenko:2014pka}.

The appearance of $\cosh$ functions and their behavior with respect to partial derivatives will prove useful in finding the partial pressures. This behavior can be summarized as follows in the case of $\mu_B$, but it applies to all chemical potentials. For even $n$, one has:
\begin{equation}
    \frac{\partial^n (P/T^4)}{\partial(\mu_B/T)^n}=F(T)\cosh\left(\frac{\mu_B}{T}\right) 
\end{equation}
and for odd $n$:
\begin{equation}
    \frac{\partial^n (P/T^4)}{\partial(\mu_B/T)^n}=F(T)\sinh\left(\frac{\mu_B}{T}\right). 
\end{equation}

This allows to easily find the susceptibilities in the Boltzmann approximation, which are central to the construction of the partial pressures. Recall that susceptibilities are defined as partial derivatives of the partition function, with respect to chemical potentials:
\begin{equation}
   \chi_{jkl}^{BSQ}=\frac{\partial^{(j+k+l)}(P/T^4)}{\partial(\mu_B/T)^j \partial(\mu_S/T)^k \partial(\mu_Q/T)^l} \Biggr|_{\Vec{\mu}=0} \, ,
\end{equation}
or, by introducing the dimensionless chemical potentials \mbox{$\hat{\mu}_X=\mu_X/T$}:
\begin{equation}
\label{susceptibilities}
   \chi_{jkl}^{BSQ} = \frac{\partial^{(j+k+l)}(P/T^4)}{{\partial\hat{\mu}_B}^j {\partial\hat{\mu}_S}^k {\partial\hat{\mu}_Q}^l} \Biggr|_{\Vec{\mu}=0}.
\end{equation}
These partial derivatives act on the full pressure which can be rewritten --- separating out the different $(B, S, Q)$ contributions--- as follows
\begin{align}
\label{fullpress}
P\left(\hat{\mu}_{B}, \hat{\mu}_{S}, \hat{\mu}_{Q}\right) 
&=P_{000} \!+\! P_{00|1|} \cosh \left(\hat{\mu}_{Q}\right) \!+\! P_{100} \cosh \left(\hat{\mu}_{B}\right) 
\nonumber \\
& \hspace{-1cm} + P_{101} \cosh \left(\hat{\mu}_{B}+\hat{\mu}_{Q}\right)+P_{10-1} \cosh \left(\hat{\mu}_{B}-\hat{\mu}_{Q}\right) 
\nonumber \\
& \hspace{-1cm} + P_{102} \cosh \left(\hat{\mu}_{B}+2 \hat{\mu}_{Q}\right)+P_{0|1| 0} \cosh \left(\hat{\mu}_{S}\right) 
\nonumber \\
& \hspace{-1cm} + P_{0|1||1|} \cosh \left(\hat{\mu}_{S}+\hat{\mu}_{Q}\right)+P_{1|1| 0} \cosh \left(\hat{\mu}_{B}-\hat{\mu}_{S}\right) 
\nonumber \\
& \hspace{-1cm} + P_{1|1| 1} \cosh \left(\hat{\mu}_{B}-\hat{\mu}_{S}+\hat{\mu}_{Q}\right) 
\nonumber \\
& \hspace{-1cm} + P_{1|1|-1} \cosh \left(\hat{\mu}_{B}-\hat{\mu}_{S}-\hat{\mu}_{Q}\right) 
\nonumber \\
& \hspace{-1cm} + P_{1|2| 0} \cosh \left(\hat{\mu}_{B}-2 \hat{\mu}_{S}\right) 
\nonumber \\
& \hspace{-1cm} + P_{1|2||1|} \cosh \left(\hat{\mu}_{B}-2 \hat{\mu}_{S}-\hat{\mu}_{Q}\right) 
\nonumber 
\nonumber \\
& \hspace{-1cm} + P_{1|3||1|} \cosh \left(\hat{\mu}_{B}-3 \hat{\mu}_{S}-\hat{\mu}_{Q}\right)
\end{align}
as was shown in \cite{Noronha-Hostler:2016rpd}.

\section{Linear Combinations for \linebreak Partial Pressure}
\label{sec:lincomb}

In this Section, we first introduce our method to isolate the contribution of different hadron families to the lattice QCD pressure, exploiting the HRG model features discussed above.
We then explain how linear combinations of lattice QCD susceptibilities are selected for each partial pressure, and we present lattice QCD results for those, in comparison to HRG model curves.

\subsection{Method}
\label{subsec:method}

By applying the definition of susceptibilities from Eq.~\eqref{susceptibilities} to the pressure as expressed in Eq.~\eqref{fullpress}, one can obtain a linear system of equations of the type:
\begin{equation}
    \chi^{BSQ}_{jkl} = \sum_{a,b,c} ({B_a}^j \times {S_b}^k \times {Q_c}^l) \, P_{abc} / T^4
\end{equation}
that relates susceptibilities to the partial pressures $P_{abc}$ introduced in Eq. (\ref{fullpress}). 
Since we want to express these 13 partial pressures in terms of susceptibilities from lattice QCD, we derive 21 equations corresponding to the whole set of second- and fourth-order susceptibilities: 
\begin{multline}
    \{\chi^B_2,\chi^S_2,\chi^Q_2,\chi^{BQ}_{11},\chi^{BS}_{11},\chi^{SQ}_{11},\chi^{B}_{4},\chi^{S}_{4}, 
    \\
    \chi^{Q}_{4},\chi^{BQ}_{31},\chi^{BS}_{31},\chi^{SQ}_{13},\chi^{BQ}_{13},\chi^{BS}_{13},\chi^{SQ}_{31},
    \\
    \chi^{BQ}_{22},\chi^{BS}_{22},\chi^{SQ}_{22},\chi^{BSQ}_{211},\chi^{BSQ}_{121},\chi^{BSQ}_{112}\} 
    \, ,
\end{multline}
all odd terms in $j+k+l$ being 0. Note that, in the convention for the susceptibilities $\chi_{ijk}^{BSQ}$, when either $i,~j$ or $k$ is zero we drop it, together with the corresponding conserved charge, from the susceptibility name.
The complete set of equations thus obtained is summarized in matrix form in Eq. \eqref{matrix_form} in Appendix \ref{sec:matrix_form}.
Since we want to obtain 13 partial pressures in terms of 21 susceptibilities, the system is clearly overdetermined \cite{Bazavov:2013dta}.

For this reason, we restrict the analysis to different subsets of susceptibilities, solving for each partial pressure individually. These subsets are chosen to minimize the uncertainty on the lattice QCD results in the hadronic phase, where e.g. higher order derivatives with respect to $\mu_Q$ lead to large error-bars due to the taste symmetry breaking in the staggered fermion formalism.
We begin by describing the linear system in matrix form:
\begin{equation}
    \mathcal{X} = M \mathcal{P}
    \label{system}
\end{equation}
where, in our case, $\mathcal{X} \in \mathbb{R}^{21}$ is the vector of susceptibilities, $\mathcal{P} \in \mathbb{R}^{13}$ is the vector of partial pressures, and $M$ is the mapping from $\mathcal{P}$ to $\mathcal{X}$ described by the coefficient matrix in Eq.~\eqref{matrix_form}.

Our goal is to express each partial pressure $P_{abc}$ as a linear combination of susceptibilities, in the form:
\begin{equation}
    \label{linsumform}
    P_{abc} = \sum_{m}^{21} \alpha_{m} \, \chi^{BSQ}_{jkl}
\end{equation}
where $\alpha_m \in \mathbb{R}$, and $m$ is the index running over the 21 $jkl$ combinations for second- and fourth-order $BSQ$ susceptibilities. These $\alpha_m$ are constants to be determined by solving the system of equations \eqref{system}. 

\subsection{Linear combination selection and results}
\label{subsec:combinations}

In this paper, we employ the latest $BSQ$ susceptibilities of order 2 and 4 \cite{jahan_2025_16749145} obtained by smoothly merging continuum estimated lattice QCD results from Ref.~\cite{Bellwied:2015lba} with HRG data at low temperature \cite{Alba:2017mqu}; more details are given in Ref.~\cite{Abuali:2025tbd}. 
This way, we ensure that the solutions converge to the HRG partial pressures at low temperature, which are calculated using the same PDG2016+ particle list \cite{SanMartin:2023zhv}.

It is important to note that the values of the susceptibilities in general are model-dependent, governed by the specific framework used to describe strongly interacting matter. 
Consequently, different models will produce significant variations in the susceptibilities, and thereby the susceptibility sets will produce different partial pressures for a given solution. The lattice QCD partial pressures obtained here can be used to benchmark hadronic models below the transition temperature.

Many solutions exhibit a similar behavior, with only small but noticeable differences, including a variation in the magnitude of the error-bars. For instance, some solutions produce large errors that provide limited insight when compared to the HRG partial pressures. Consequently, we prioritize solutions with smaller error values.

Notice that the different linear combinations corresponding to the same partial pressure agree in the hadronic phase within error-bar, as they all yield the contribution of the same hadronic family to the full QCD pressure. Above the phase transition, however, this is no longer true: the different linear combinations lose their partial pressure meaning, and in general do not agree with each other in the quark phase.

\begin{figure}
    \centering
    \includegraphics[width=0.8\linewidth]{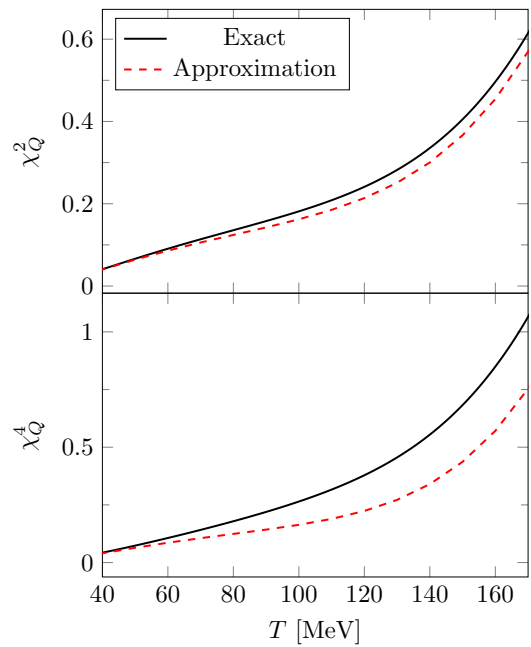}
    \caption{Values of $\chi^2_Q$ and $\chi^4_Q$ from the HRG model as a function of temperature, contrasting the values derived from the full HRG pressure (in full black lines) and from the Boltzmann-approximated pressure (in dashed red line).}
    \label{fig:Chi4Comp}
\end{figure}

An important step in obtaining the pressure decomposition in Eq. \eqref{fullpress} is the application of the Boltzmann approximation to the partition function. The numerical values of the pressure obtained from the full and approximated expressions show excellent agreement across the entire temperature range. When considering higher-order derivatives to calculate the susceptibilities, we find that all but two of them, namely $\chi^2_Q$ and $\chi^4_Q$, exhibit good agreement between the full and approximated results.

The numerical difference in  $\chi^2_Q$ and $\chi^4_Q$, shown in Fig.~\ref{fig:Chi4Comp}, plays an important role in the methods to derive the partial pressure, whether analytical or using the susceptibilities. 
If all the susceptibilities are treated within the Boltzmann approximation, each solution from the linear system reproduces the partial pressure obtained analytically through partial particle list summation in the HRG model. 
However, due the discrepancies in $\chi^2_Q$ and $\chi^4_Q$, solutions based on the full HRG model susceptibilities (obtained by deriving the full HRG pressure with respect to the electric charge chemical potential) or lattice QCD will fail to accurately reconstruct the partial pressures below the transition temperature, whenever either of these two susceptibilities is included in the calculation.
This is likely due to the fact that the electric sector receives its main contribution from pions. The small pion mass $m_\pi$ limits the use of the Boltzmann approximation, which relies on the criterion that $(m_i - \mu) \geq T$ for hadron $i$.

In the following, we show the equations for our choice of susceptibility combination for each of the thirteen partial pressures $P_{abc}$, together with plots showing their behavior as functions of the temperature, compared with HRG model partial pressures.
Note that these combinations are encoded in the associated \textit{Partial Pressures} module~\cite{jahan_2026_19747641} of the MUSES Calculation Engine~\cite{calculation_engine_all_versions}, which allows for anyone to compute these partial pressures given a set of second- and fourth-order susceptibilities provided by the user.

\begin{figure}[!h]
    \centering
    \includegraphics[scale=1]{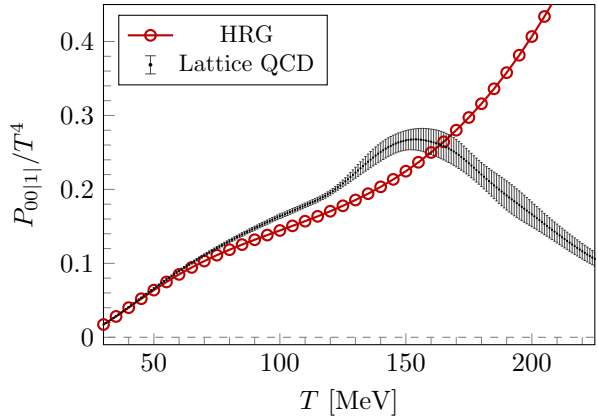}
    \vspace*{-10pt}
    \caption{$P_{00|1|}/T^4$ obtained from the linear combination of lattice susceptibilities (black) given in Eq.~\eqref{eq:p001}, compared with HRG calculations (red), as a function of temperature.}
    \label{fig:p001}
\end{figure}
\vspace*{-10pt}
\begin{equation}
    \label{eq:p001}
    \begin{split}
        P_{00|1|} / T^4 &= \chi^{Q}_{2} + 2 \chi^{BS}_{11} + \chi^{BS}_{13} + \chi^{SQ}_{31} + 3 \chi^{BS}_{22} \\
        &\hspace{12pt} - \chi^{BQ}_{22} - 2 \chi^{SQ}_{22} + \chi^{BSQ}_{121} - 2 \chi^{BSQ}_{112}
    \end{split} 
\end{equation}

\begin{figure}[!h]
    \includegraphics[scale=1]{plots/P100.png}
    \caption{Same as Fig.~\ref{fig:p001} for $P_{100}/T^4$ taken  from Eq.~\eqref{eq:p100}.}
    \label{fig:p100}
\end{figure}
\begin{equation}
    \label{eq:p100}
    \begin{split}
        P_{100} / T^4 &= \frac{1}{24} \chi^{S}_{2} - \frac{1}{2} \chi^{BQ}_{11} + \frac{1}{2} \chi^{SQ}_{11} + \chi^{B}_{4} - \frac{1}{24} \chi^{S}_{4}  \\
        &\hspace{12pt}  - \frac{1}{4} \chi^{BS}_{13} + \frac{1}{2} \chi^{BQ}_{13} + \frac{5}{4} \chi^{BS}_{31} - \chi^{BQ}_{22}  \\
        &\hspace{12pt} - \frac{1}{2} \chi^{SQ}_{22} - \frac{1}{2} \chi^{BSQ}_{211} - \frac{3}{2} \chi^{BSQ}_{112}
    \end{split}
\end{equation}

\begin{figure}[!h]
    \includegraphics[scale=1]{plots/P101.png}
    \caption{Same as Fig.~\ref{fig:p001} for $P_{101}/T^4$ taken  from Eq.~\eqref{eq:p101}.}
    \label{fig:p101}
\end{figure}
\begin{equation}
    \label{eq:p101}
    \begin{split}
        P_{101} / T^4  &= \frac{1}{24} \chi^{S}_{2} + \frac{1}{12} \chi^{BS}_{11} + \chi^{BQ}_{11} - \frac{1}{24} \chi^{S}_{4}  \\
        &\hspace{12pt} - \frac{1}{12} \chi^{BS}_{13}  - \frac{1}{2} \chi^{BQ}_{13} + \frac{1}{3} \chi^{SQ}_{13} + \frac{1}{6} \chi^{SQ}_{31}  \\
        &\hspace{12pt} + \frac{1}{2} \chi^{BQ}_{22} - \frac{1}{2} \chi^{SQ}_{22}
    \end{split}
\end{equation}

\pagebreak

\begin{figure}[!h]
    \includegraphics[scale=1]{plots/P10-1.png}
    \caption{Same as Fig.~\ref{fig:p001} for $P_{10-1}/T^4$ taken  from Eq.~\eqref{eq:p10-1}.}
    \label{fig:p10-1}
\end{figure}
\begin{equation}
    \label{eq:p10-1}
    \begin{split}
        P_{10-1} / T^4  &= \frac{1}{8} \chi^{S}_{2} + \frac{1}{4} \chi^{BS}_{11} - \frac{1}{8} \chi^{S}_{4} - \frac{1}{4} \chi^{BS}_{13}  \\
        &\hspace{12pt} - \frac{1}{6} \chi^{BQ}_{13} + \frac{1}{6} \chi^{SQ}_{13} - \frac{1}{3} \chi^{BQ}_{31}  \\
        &\hspace{12pt} + \frac{1}{3} \chi^{SQ}_{31} + \frac{1}{2} \chi^{BQ}_{22} - \frac{1}{2} \chi^{SQ}_{22} - \chi^{BSQ}_{211}
    \end{split}
\end{equation}

\begin{figure}[!h]
    \includegraphics[scale=1]{plots/P102.png}
    \caption{Same as Fig.~\ref{fig:p001} for $P_{102}/T^4$ taken  from Eq.~\eqref{eq:p102}.}
    \label{fig:p102}
\end{figure}
\begin{equation}
    \label{eq:p102}
    \begin{split}
        P_{102} / T^4  &= \frac{1}{6} \chi^{BQ}_{13} - \frac{1}{6} \chi^{BQ}_{31} 
    \end{split}
\end{equation}

\pagebreak

\begin{figure}[!h]
    \includegraphics[scale=1]{plots/P010.png}
    \caption{Same as Fig.~\ref{fig:p001} for $P_{0|1|0}/T^4$ taken  from Eq.~\eqref{eq:p010}.}
    \label{fig:p010}
\end{figure}
\begin{equation}
    \label{eq:p010}
    \begin{split}
        P_{0|1|0} / T^4 &= \frac{2}{3} \chi^{S}_{2} - \chi^{SQ}_{11} - \frac{1}{3} \chi^{S}_{4} + \chi^{BS}_{13}    \\
        &\hspace{12pt} - \chi^{SQ}_{31} + \chi^{SQ}_{22} - 2 \chi^{BSQ}_{121} + \chi^{BSQ}_{112}
    \end{split}
\end{equation}

\begin{figure}[!h]
    \includegraphics[scale=1]{plots/P011.png}
    \caption{Same as Fig.~\ref{fig:p001} for $P_{0|1||1|}/T^4$ taken  from Eq.~\eqref{eq:p011}.}
    \label{fig:p011}
\end{figure}
\begin{equation}
    \label{eq:p011}
    \begin{split}
        P_{0|1||1|} / T^4  &= \frac{2}{3} \chi^{SQ}_{11} + \frac{1}{3} \chi^{BS}_{13} + \frac{2}{3} \chi^{BS}_{31} \\
        &\hspace{12pt} + \frac{1}{3} \chi^{SQ}_{31}  +  \chi^{BS}_{22}+ \chi^{BSQ}_{121}
    \end{split}
\end{equation}

\pagebreak

\begin{figure}[!h]
    \includegraphics[scale=1]{plots/P110.png}
    \caption{Same as Fig.~\ref{fig:p001} for $P_{1|1|0}/T^4$ taken  from Eq.~\eqref{eq:p110}.}
    \label{fig:p110}
\end{figure}
\begin{equation}
    \label{eq:p110}
    \begin{split}
        P_{1|1|0} / T^4  &= \frac{1}{6} \chi^{S}_{2} - \chi^{BS}_{11} - \chi^{SQ}_{11} - \frac{1}{6} \chi^{S}_{4}  \\
        &\hspace{12pt}  + \chi^{SQ}_{31} + 2 \chi^{BSQ}_{211} + 2 \chi^{BSQ}_{121} + \chi^{BSQ}_{112}
    \end{split}
\end{equation}

\begin{figure}[!h]
    \includegraphics[scale=1]{plots/P111.png}
    \caption{Same as Fig.~\ref{fig:p001} for $P_{1|1|1}/T^4$ taken  from Eq.~\eqref{eq:p111}.}
    \label{fig:p111}
\end{figure}
\begin{equation}
    \label{eq:p111}
    \begin{split}
        P_{1|1|1} / T^4  &= -\frac{1}{2} \chi^{BSQ}_{211} -\frac{1}{2} \chi^{BSQ}_{112}
    \end{split}
\end{equation}

\pagebreak

\begin{figure}[!h]
    \includegraphics[scale=1]{plots/P11-1.png}
    \caption{Same as Fig.~\ref{fig:p001} for $P_{1|1|-1}/T^4$ taken  from Eq.~\eqref{eq:p11-1}.}
    \label{fig:p11-1}
\end{figure}
\begin{equation}
    \label{eq:p11-1}
    \begin{split}
        P_{1|1|-1} / T^4  &= -\frac{1}{12} \chi^{S}_{2} - \frac{1}{2} \chi^{BS}_{11} + \frac{1}{12} \chi^{S}_{4} - \frac{1}{2} \chi^{SQ}_{13}  \\
        &\hspace{12pt} - \frac{1}{2} \chi^{BS}_{22} + \frac{1}{2} \chi^{SQ}_{22} + \frac{5}{2} \chi^{BSQ}_{211} + \frac{3}{2} \chi^{BSQ}_{121} 
    \end{split}
\end{equation}

\begin{figure}[!h]
    \includegraphics[scale=1]{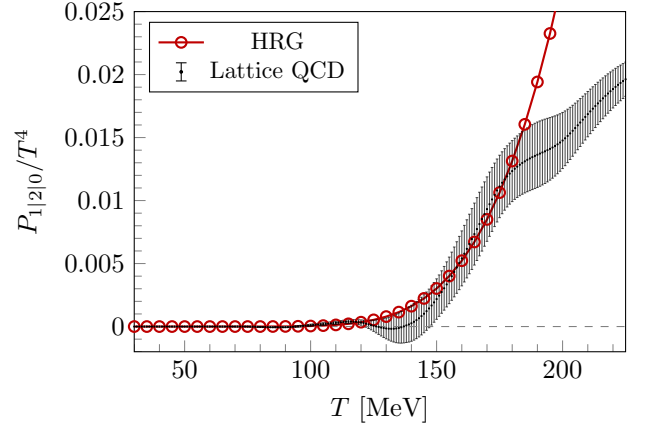}
    \caption{Same as Fig.~\ref{fig:p001} for $P_{1|2|0}/T^4$ taken  from Eq.~\eqref{eq:p120}.}
    \label{fig:p120}
\end{figure}
\begin{equation}
    \label{eq:p120}
    \begin{split}
        P_{1|2|0} / T^4  &= \frac{1}{36} \chi^{S}_{2} - \frac{1}{36} \chi^{S}_{4} - \frac{1}{3} \chi^{BS}_{13}  \\
        &\hspace{12pt} + \frac{1}{6} \chi^{SQ}_{13} - \frac{1}{6} \chi^{SQ}_{31} - \frac{1}{3} \chi^{BS}_{22}
    \end{split}
\end{equation}

\pagebreak

\begin{figure}[!h]
    \includegraphics[scale=1]{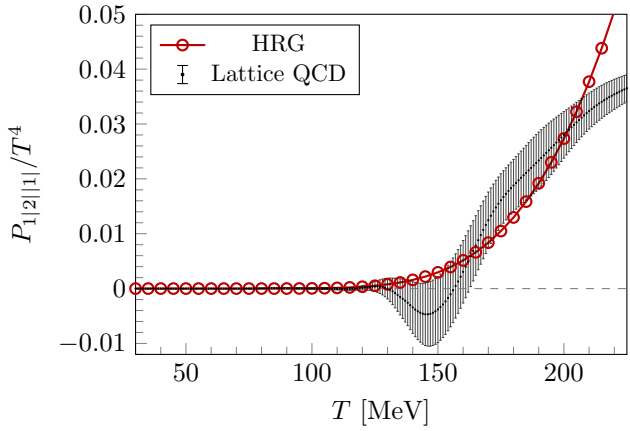}
    \caption{Same as Fig.~\ref{fig:p001} for $P_{1|2||1|}/T^4$ taken  from Eq.~\eqref{eq:p121}.}
    \label{fig:p121}
\end{figure}
\begin{equation}
    \label{eq:p121}
    \begin{split}
        P_{1|2||1|} / T^4  &= \frac{1}{9} \chi^{S}_{2} - \frac{1}{6} \chi^{SQ}_{11} - \frac{1}{9} \chi^{S}_{4} + \frac{2}{3} \chi^{BS}_{31}  \\
        &\hspace{12pt} + \frac{1}{6} \chi^{SQ}_{31} + \frac{2}{3} \chi^{BS}_{22}
    \end{split}
\end{equation}

\begin{figure}[!h]
    \includegraphics[scale=1]{plots/P131.png}
    \caption{Same as Fig.~\ref{fig:p001} for $P_{1|3||1|}/T^4$ taken  from Eq.~\eqref{eq:p131}.}
    \label{fig:p131}
\end{figure}
\label{last}
\begin{equation}
    \label{eq:p131}
    \begin{split}
        P_{1|3||1|} / T^4  &= \frac{1}{12} \chi^{SQ}_{13} + \frac{1}{6} \chi^{SQ}_{31} - \frac{1}{4} \chi^{SQ}_{22} \\
        &\hspace{12pt} + \frac{1}{4} \chi^{BSQ}_{121} - \frac{1}{4} \chi^{BSQ}_{112}
    \end{split}
\end{equation}

As shown in the above Figures, the partial pressure linear combinations selected using lattice QCD susceptibilities are almost all in good agreement with the HRG-calculated partial pressures. 
Notice that the HRG partial pressures have been obtained with two different methods: the dots indicate the result from the linear combination of susceptibilities, while the full lines show the partial pressures obtained by summing only over the selected hadronic family in Eq. \eqref{pp}. This serves as a proof that our selected linear combinations correspond indeed to the desired partial pressure.
The disagreement shown by $P_{001}$ below the transition temperature $T_0$ is due to the major difference with the Boltzmann approximation for $\chi^2_Q$ and $\chi^4_Q$. For this partial pressure, every linear combination contained at least one of these susceptibilities.

Only $P_{131}$ clearly overestimates the HRG partial pressure, which arises from the $\Omega$ hyperon family. This discrepancy could be explained by the fact that there is still little known about the resonance spectroscopy of this hadron species \cite{Crede:2024hur}. 
The incomplete particle list we use would then naturally lead to a lower partial pressure than expected, as corroborated by the study from Ref.~\cite{Alba:2017mqu}.
A similar effect, although less pronounced, can be observed for $P_{121}$ too, which corresponds to the $\Xi$ hyperon family.

\section{Conclusions}
\label{sec:concl}

Under the Boltzmann approximation of the HRG model partition function, one can represent the hadronic pressure in terms of partial pressures, where different hadron families are grouped according to their baryon number $B$, strangeness $S$ and electric charge $Q$. Applying the definition of susceptibilities to this pressure representation, we obtained a linear system of the susceptibilities up to fourth order in terms of partial pressures.

With 21 susceptibilities to consider, and 13 partial pressures, the system is clearly overdetermined. For this reason, we used the minimization of the lattice QCD error-bars as a criterion for picking our linear combinations, which are then benchmarked against HRG model results obtained summing the total pressure of the system only over the targeted hadronic family.
The lattice QCD results obtained here are based on state-of-the-art continuum estimates for the $BSQ$ susceptibilities up to fourth order, and can be used to test hadronic models in order to costrain their parameters.
The comparison between lattice QCD results for the different partial pressures and the HRG model curves generally shows good agreement, with the exception of the baryons containing three strange quarks, for which the HRG model result underestimates the lattice QCD curve. This might be due to missing resonances in this sector from the PDG2016+ list utilized here. 
Future extensions of this analysis can include hadrons carrying charm $C$, when a full set of $BSQC$ susceptibilities becomes available.

\section*{Acknowledgements}
This material is based upon work supported by the National Science Foundation under grants No. PHY-2208724, PHY-2116686 and PHY-2514763, and within the framework of the MUSES collaboration, under Grant No. OAC-2103680. This material is also based upon work supported by the U.S. Department of Energy, Office of Science, Office of Nuclear Physics, under Award Number DE-SC0022023, as well as by the National Aeronautics and Space Agency (NASA) under Award Number 80NSSC24K0767.

\appendix

\setcounter{MaxMatrixCols}{13}
    
\section{Matrix equation form between partial pressures and susceptibilities}\label{sec:matrix_form}

By using the expression of pressure given in Eq.~\eqref{fullpress} into the definition of susceptibilities from Eq.~\eqref{susceptibilities}, we obtain a linear combination of partial pressures for each susceptibility, up to fourth-order. We represent the relationship of these susceptibilities with the partial pressures through the following matrix equation:

\begin{widetext}
    \begin{equation}\label{matrix_form}
        T^4 {\renewcommand{\arraystretch}{1.2}
        \begin{bmatrix}
            \chi^{B}_{2}  \\
            \chi^{S}_{2}  \\
            \chi^{Q}_{2}  \\
            \chi^{BS}_{11}  \\
            \chi^{BQ}_{11}  \\
            \chi^{SQ}_{11}  \\
            \chi^{B}_{4}  \\
            \chi^{S}_{4}  \\
            \chi^{Q}_{4}  \\
            \chi^{BS}_{13}  \\
            \chi^{BQ}_{13}  \\
            \chi^{SQ}_{13}  \\
            \chi^{BS}_{31}  \\
            \chi^{BQ}_{31}  \\
            \chi^{SQ}_{31}  \\
            \chi^{BS}_{22}  \\
            \chi^{BQ}_{22}  \\
            \chi^{SQ}_{22}  \\
            \chi^{BSQ}_{211}  \\
            \chi^{BSQ}_{121}  \\
            \chi^{BSQ}_{112}  \\
        \end{bmatrix}
        }
        =   
        {\renewcommand{\arraystretch}{1.2}
        \begin{bmatrix}
            0 &  1  &  1  &  1  &  1  &  0  &  0  &  1  &  1  &  1  &  1  &  1  &  1  \\
            0  &  0  &  0  &  0  &  0  &  1  &  1  &  1  &  1  &  1  &  4  &  4  &  9  \\
            1  &  0  &  1  &  1  &  4  &  0  &  1  &  0  &  1  &  1  &  0  &  1  &  1  \\
            0  &  0  &  0  &  0  &  0  &  0  &  0  &  -1  &  -1  &  -1  &  -2  &  -2  &  -3  \\
            0  &  0  &  1  &  -1  &  2  &  0  &  0  &  0  &  1  &  -1  &  0  &  -1  &  -1  \\
            0  &  0  &  0  &  0  &  0  &  0  &  1  &  0  &  -1  &  1  &  0  &  2  &  3  \\
            0  &  1  &  1  &  1  &  1  &  0  &  0  &  1  &  1  &  1  &  1  &  1  &  1  \\
            0  &  0  &  0  &  0  &  0  &  1  &  1  &  1  &  1  &  1  &  16  &  16  &  81  \\
            1  &  0  &  1  &  1  &  16  &  0  &  1  &  0  &  1  &  1  &  0  &  1  &  1  \\
            0  &  0  &  0  &  0  &  0  &  0  &  0  &  -1  &  -1  &  -1  &  -8  &  -8  &  -27  \\
            0  &  0  &  1  &  -1  &  8  &  0  &  0  &  0  &  1  &  -1  &  0  &  -1  &  -1  \\
            0  &  0  &  0  &  0  &  0  &  0  &  1  &  0  &  -1  &  1  &  0  &  2  &  3  \\
            0  &  0  &  0  &  0  &  0  &  0  &  0  &  -1  &  -1  &  -1  &  -2  &  -2  &  -3  \\
            0  &  0  &  1  &  -1  &  2  &  0  &  0  &  0  &  1  &  -1  &  0  &  -1  &  -1  \\
            0  &  0  &  0  &  0  &  0  &  0  &  1  &  0  &  -1  &  1  &  0  &  8  &  27  \\
            0  &  0  &  0  &  0  &  0  &  0  &  0  &  1  &  1  &  1  &  4  &  4  &  9  \\
            0  &  0  &  1  &  1  &  4  &  0  &  0  &  0  &  1  &  1  &  0  &  1  &  1  \\
            0  &  0  &  0  &  0  &  0  &  0  &  1  &  0  &  1  &  1  &  0  &  4  &  9  \\
            0  &  0  &  0  &  0  &  0  &  0  &  0  &  0  &  -1  &  1  &  0  &  2  &  3  \\
            0  &  0  &  0  &  0  &  0  &  0  &  0  &  0  &  1  &  -1  &  0  &  -4  &  -9  \\
            0  &  0  &  0  &  0  &  0  &  0  &  0  &  0  &  -1  &  -1  &  0  &  -2  &  -3  \\
        \end{bmatrix}
        }
        \begin{bmatrix}
            P_{001}  \\
            P_{100}  \\
            P_{101}  \\
            P_{10-1}  \\
            P_{102}  \\
            P_{010}  \\
            P_{011}  \\
            P_{110}  \\
            P_{111}  \\
            P_{11-1}  \\
            P_{120}  \\
            P_{121}  \\
            P_{131}  \\
        \end{bmatrix}_{BSQ}
    \end{equation}
\end{widetext}

The above formula represents a system of equations where 21 susceptibilities can be expressed in terms of 13 partial pressures. The inversion of this system yields the desired partial pressure as a linear combination of susceptibilities. Due to the disparity in the number of partial pressures and susceptibilities, the system is overdetermined, and we need a criterion to pick a linear combination of susceptibilities to express a partial pressure. As explained in the main text, we choose to minimize the uncertainty in the lattice QCD results, so that they can be more useful in constraining hadronic model predictions for the same quantities.

\bibliography{all}

@article{Abuali:2025tbd,
    author = "Abuali, Ahmed and Bors{\'a}nyi, Szabolcs and Fodor, Zolt{\'a}n and Jahan, Johannes and Kahangirwe, Micheal and Parotto, Paolo and P{\'a}sztor, Attila and Ratti, Claudia and Shah, Hitansh and Trabulsi, Seth A.",
    title = "{New 4D lattice QCD equation of state: Extended density coverage from a generalized T' expansion}",
    eprint = "2504.01881",
    archivePrefix = "arXiv",
    primaryClass = "hep-lat",
    doi = "10.1103/2dmh-26yh",
    journal = "Phys. Rev. D",
    volume = "112",
    number = "5",
    pages = "054502",
    year = "2025"
}

@article{Alba:2017mqu,
    author = "Alba, Paolo and others",
    title = "{Constraining the hadronic spectrum through QCD thermodynamics on the lattice}",
    eprint = "1702.01113",
    archivePrefix = "arXiv",
    primaryClass = "hep-lat",
    doi = "10.1103/PhysRevD.96.034517",
    journal = "Phys. Rev. D",
    volume = "96",
    number = "3",
    pages = "034517",
    year = "2017"
}

@article{Bazavov:2013dta,
    author = "Bazavov, A. and others",
    title = "{Strangeness at high temperatures: from hadrons to quarks}",
    eprint = "1304.7220",
    archivePrefix = "arXiv",
    primaryClass = "hep-lat",
    doi = "10.1103/PhysRevLett.111.082301",
    journal = "Phys. Rev. Lett.",
    volume = "111",
    pages = "082301",
    year = "2013"
}

@article{ALICE:2022wpn,
    author = "Acharya, Shreyasi and others",
    collaboration = "ALICE",
    title = "{The ALICE experiment: a journey through QCD}",
    eprint = "2211.04384",
    archivePrefix = "arXiv",
    primaryClass = "nucl-ex",
    reportNumber = "CERN-EP-2022-227",
    doi = "10.1140/epjc/s10052-024-12935-y",
    journal = "Eur. Phys. J. C",
    volume = "84",
    number = "8",
    pages = "813",
    year = "2024"
}

@article{Annala:2017llu,
    author = "Annala, Eemeli and Gorda, Tyler and Kurkela, Aleksi and Vuorinen, Aleksi",
    title = "{Gravitational-wave constraints on the neutron-star-matter Equation of State}",
    eprint = "1711.02644",
    archivePrefix = "arXiv",
    primaryClass = "astro-ph.HE",
    reportNumber = "CERN-TH-2017-236",
    doi = "10.1103/PhysRevLett.120.172703",
    journal = "Phys. Rev. Lett.",
    volume = "120",
    number = "17",
    pages = "172703",
    year = "2018"
}

@article{Aoki:2006we,
    author = "Aoki, Y. and Endrodi, G. and Fodor, Z. and Katz, S. D. and Szabo, K. K.",
    title = "{The Order of the quantum chromodynamics transition predicted by the standard model of particle physics}",
    eprint = "hep-lat/0611014",
    archivePrefix = "arXiv",
    doi = "10.1038/nature05120",
    journal = "Nature",
    volume = "443",
    pages = "675--678",
    year = "2006"
}

@article{Bellwied:2015lba,
    author = "Bellwied, R. and Borsanyi, S. and Fodor, Z. and Katz, S. D. and Pasztor, A. and Ratti, C. and Szabo, K. K.",
    title = "{Fluctuations and correlations in high temperature QCD}",
    eprint = "1507.04627",
    archivePrefix = "arXiv",
    primaryClass = "hep-lat",
    doi = "10.1103/PhysRevD.92.114505",
    journal = "Phys. Rev. D",
    volume = "92",
    number = "11",
    pages = "114505",
    year = "2015"
}

@article{Bzdak:2019pkr,
    author = "Bzdak, Adam and Esumi, Shinichi and Koch, Volker and Liao, Jinfeng and Stephanov, Mikhail and Xu, Nu",
    title = "{Mapping the Phases of Quantum Chromodynamics with Beam Energy Scan}",
    eprint = "1906.00936",
    archivePrefix = "arXiv",
    primaryClass = "nucl-th",
    doi = "10.1016/j.physrep.2020.01.005",
    journal = "Phys. Rept.",
    volume = "853",
    pages = "1--87",
    year = "2020"
}

@article{CABIBBO197567,
    author = {N. Cabibbo and G. Parisi},
    title = {Exponential hadronic spectrum and quark liberation},
    journal = {Physics Letters B},
    volume = {59},
    number = {1},
    pages = {67-69},
    year = {1975},
    issn = {0370-2693},
    doi = {https://doi.org/10.1016/0370-2693(75)90158-6},
    url = {https://www.sciencedirect.com/science/article/pii/0370269375901586},
}

@article{Crede:2024hur,
    author = "Crede, Volker and Yelton, John",
    title = "{70 years of hyperon spectroscopy: a review of strange $\Xi$, $\Omega$ baryons, and the spectrum of charmed and bottom baryons}",
    eprint = "2502.08815",
    archivePrefix = "arXiv",
    primaryClass = "hep-ex",
    doi = "10.1088/1361-6633/ad7610",
    journal = "Rept. Prog. Phys.",
    volume = "87",
    number = "10",
    pages = "106301",
    year = "2024"
}

@article{deForcrand:2002hgr,
    author = "de Forcrand, Philippe and Philipsen, Owe",
    title = "{The QCD phase diagram for small densities from imaginary chemical potential}",
    eprint = "hep-lat/0205016",
    archivePrefix = "arXiv",
    reportNumber = "MIT-CTP-3270, CERN-TH-2002-102",
    doi = "10.1016/S0550-3213(02)00626-0",
    journal = "Nucl. Phys. B",
    volume = "642",
    pages = "290--306",
    year = "2002"
}

@article{DElia:2002tig,
    author = "D'Elia, Massimo and Lombardo, Maria-Paola",
    title = "{Finite density QCD via imaginary chemical potential}",
    eprint = "hep-lat/0209146",
    archivePrefix = "arXiv",
    reportNumber = "GEF-TH-2002-12",
    doi = "10.1103/PhysRevD.67.014505",
    journal = "Phys. Rev. D",
    volume = "67",
    pages = "014505",
    year = "2003"
}

@article{Du:2024wjm,
    author = "Du, Lipei and Sorensen, Agnieszka and Stephanov, Mikhail",
    title = "{The QCD phase diagram and Beam Energy Scan physics: A theory overview}",
    eprint = "2402.10183",
    archivePrefix = "arXiv",
    primaryClass = "nucl-th",
    reportNumber = "INT-PUB-24-017",
    doi = "10.1142/9789811294679_0007",
    journal = "Int. J. Mod. Phys. E",
    volume = "33",
    number = "07",
    pages = "2430008",
    year = "2024"
}

@article{Fischer:2009wc,
    author = "Fischer, Christian S.",
    title = "{Deconfinement phase transition and the quark condensate}",
    eprint = "0904.2700",
    archivePrefix = "arXiv",
    primaryClass = "hep-ph",
    doi = "10.1103/PhysRevLett.103.052003",
    journal = "Phys. Rev. Lett.",
    volume = "103",
    pages = "052003",
    year = "2009"
}

@article{Gorenstein:1981fa,
    author = "Gorenstein, Mark I. and Petrov, V. K. and Zinovev, G. M.",
    title = "{Phase Transition in the Hadron Gas Model}",
    reportNumber = "ITF-81-74E",
    doi = "10.1016/0370-2693(81)90546-3",
    journal = "Phys. Lett. B",
    volume = "106",
    pages = "327--330",
    year = "1981"
}

@article{Hagedorn:1965st,
    author = "Hagedorn, R.",
    title = "{Statistical thermodynamics of strong interactions at high-energies}",
    reportNumber = "CERN-TH-520",
    journal = "Nuovo Cim. Suppl.",
    volume = "3",
    pages = "147--186",
    year = "1965"
}

@article{Kapusta:1982qd,
    author = "Kapusta, Joseph I. and Olive, Keith A.",
    title = "{Thermodynamics of Hadrons: Delimiting the Temperature}",
    reportNumber = "CERN-TH-3421",
    doi = "10.1016/0375-9474(83)90241-5",
    journal = "Nucl. Phys. A",
    volume = "408",
    pages = "478--494",
    year = "1983"
}

@article{Karsch:2001cy,
    author = "Karsch, F.",
    editor = "Plessas, Willibald and Mathelitsch, L.",
    title = "{Lattice QCD at high temperature and density}",
    eprint = "hep-lat/0106019",
    archivePrefix = "arXiv",
    reportNumber = "BI-TP-2001-10",
    doi = "10.1007/3-540-45792-5_6",
    journal = "Lect. Notes Phys.",
    volume = "583",
    pages = "209--249",
    year = "2002"
}

@article{Kogut:1974ag,
    author = "Kogut, John B. and Susskind, Leonard",
    title = "{Hamiltonian Formulation of Wilson's Lattice Gauge Theories}",
    reportNumber = "Print-74-1186 (CORNELL)",
    doi = "10.1103/PhysRevD.11.395",
    journal = "Phys. Rev. D",
    volume = "11",
    pages = "395--408",
    year = "1975"
}

@article{LIGOScientific:2018cki,
    author = "Abbott, B. P. and others",
    collaboration = "LIGO Scientific, Virgo",
    title = "{GW170817: Measurements of neutron star radii and equation of state}",
    eprint = "1805.11581",
    archivePrefix = "arXiv",
    primaryClass = "gr-qc",
    reportNumber = "LIGO-P1800115",
    doi = "10.1103/PhysRevLett.121.161101",
    journal = "Phys. Rev. Lett.",
    volume = "121",
    number = "16",
    pages = "161101",
    year = "2018"
}

@article{LIGOScientific:2020aai,
    author = "Abbott, B. P. and others",
    collaboration = "LIGO Scientific, Virgo",
    title = "{GW190425: Observation of a Compact Binary Coalescence with Total Mass $\sim 3.4 M_{\odot}$}",
    eprint = "2001.01761",
    archivePrefix = "arXiv",
    primaryClass = "astro-ph.HE",
    reportNumber = "LIGO-P190425",
    doi = "10.3847/2041-8213/ab75f5",
    journal = "Astrophys. J. Lett.",
    volume = "892",
    number = "1",
    pages = "L3",
    year = "2020"
}

@article{Noronha-Hostler:2016rpd,
    author = "Noronha-Hostler, Jacquelyn and Bellwied, Rene and Gunther, Jana and Parotto, Paolo and Pasztor, Attila and Portillo Vazquez, Israel and Ratti, Claudia",
    title = "{Kaon fluctuations from lattice QCD}",
    eprint = "1607.02527",
    archivePrefix = "arXiv",
    primaryClass = "hep-ph",
    year = "arXiv:1607.02527, 2016",
    journal = ""
}

@article{ParticleDataGroup:2024cfk,
    author = "Navas, S. and others",
    collaboration = "Particle Data Group",
    title = "{Review of particle physics}",
    doi = "10.1103/PhysRevD.110.030001",
    journal = "Phys. Rev. D",
    volume = "110",
    number = "3",
    pages = "030001",
    year = "2024"
}

@article{Ratti:2018ksb,
    author = "Ratti, Claudia",
    title = "{Lattice QCD and heavy ion collisions: a review of recent progress}",
    eprint = "1804.07810",
    archivePrefix = "arXiv",
    primaryClass = "hep-lat",
    doi = "10.1088/1361-6633/aabb97",
    journal = "Rept. Prog. Phys.",
    volume = "81",
    number = "8",
    pages = "084301",
    year = "2018"
}

@article{Ratti:2022qgf,
    author = "Ratti, Claudia",
    title = "{Equation of state for QCD from lattice simulations}",
    doi = "10.1016/j.ppnp.2022.104007",
    journal = "Prog. Part. Nucl. Phys.",
    volume = "129",
    pages = "104007",
    year = "2023"
}

@article{SanMartin:2023zhv,
    author = "San Martin, Jordi Salinas and Hirayama, Renan and Hammelmann, Jan and Karthein, Jamie M. and Parotto, Paolo and Noronha-Hostler, Jacquelyn and Ratti, Claudia and Elfner, Hannah",
    title = "{Thermodynamics of an updated hadronic resonance list and influence on hadronic transport}",
    eprint = "2309.01737",
    archivePrefix = "arXiv",
    primaryClass = "nucl-th",
    month = "9",
    year = "2023"
}

@article{Sorensen:2023zkk,
    author = "Sorensen, Agnieszka and others",
    title = "{Dense nuclear matter equation of state from heavy-ion collisions}",
    eprint = "2301.13253",
    archivePrefix = "arXiv",
    primaryClass = "nucl-th",
    reportNumber = "INT-PUB-23-001, LA-UR-23-20514, LLNL-TR-844629",
    doi = "10.1016/j.ppnp.2023.104080",
    journal = "Prog. Part. Nucl. Phys.",
    volume = "134",
    pages = "104080",
    year = "2024"
}

@article{STAR:2010vob,
    author = "Aggarwal, M. M. and others",
    collaboration = "STAR",
    title = "{An Experimental Exploration of the QCD Phase Diagram: The Search for the Critical Point and the Onset of De-confinement}",
    eprint = "1007.2613",
    archivePrefix = "arXiv",
    primaryClass = "nucl-ex",
    month = "7",
    year = "2010"
}

@article{Susskind:1976jm,
    author = "Susskind, Leonard",
    title = "{Lattice Fermions}",
    reportNumber = "PTENS-76-1",
    doi = "10.1103/PhysRevD.16.3031",
    journal = "Phys. Rev. D",
    volume = "16",
    pages = "3031--3039",
    year = "1977"
}

@article{Vovchenko:2014pka,
    author = "Vovchenko, V. and Anchishkin, D. V. and Gorenstein, M. I.",
    title = "{Hadron Resonance Gas Equation of State from Lattice QCD}",
    eprint = "1412.5478",
    archivePrefix = "arXiv",
    primaryClass = "nucl-th",
    doi = "10.1103/PhysRevC.91.024905",
    journal = "Phys. Rev. C",
    volume = "91",
    number = "2",
    pages = "024905",
    year = "2015"
}

@article{Wilson:1974sk,
    author = "Wilson, Kenneth G.",
    editor = "Taylor, J. C.",
    title = "{Confinement of Quarks}",
    reportNumber = "CLNS-262",
    doi = "10.1103/PhysRevD.10.2445",
    journal = "Phys. Rev. D",
    volume = "10",
    pages = "2445--2459",
    year = "1974"
}

@dataset{jahan_2025_16749145,
  author       = {Jahan, Johannès and
                  Parotto, Paolo},
  title        = {Continuum-estimated lattice susceptibilites of
                   order 2 and 4},
  month        = aug,
  year         = 2025,
  publisher    = {Zenodo},
  version      = {2.0.0},
  doi          = {10.5281/zenodo.16749145},
  url          = {https://doi.org/10.5281/zenodo.16749145}
}

@software{jahan_2026_19747641,
  author       = {Jahan, Johannes and
                  Gonzalez, Jonathan and
                  Nava Acuña, Angel R.},
  title        = {Partial Pressures module},
  month        = apr,
  year         = 2026,
  publisher    = {Zenodo},
  version      = {1.0.0},
  doi          = {10.5281/zenodo.19747641},
  url          = {https://doi.org/10.5281/zenodo.19747641},
}

@software{calculation_engine_all_versions,
  author       = {Manning, T. Andrew},
  title        = {{MUSES Calculation Engine}},
  month        = jan,
  year         = 2025,
  publisher    = {Zenodo},
  version      = {latest},
  doi          = {10.5281/zenodo.14721911},
  url          = {https://doi.org/10.5281/zenodo.14721911},
}

\end{document}